\documentclass[11pt,letterpaper]{article}
\usepackage[letterpaper]{geometry}
\usepackage{acl2012}
\usepackage{times}
\usepackage{latexsym}
\usepackage{amsmath}
\usepackage{multirow}
\usepackage{url}
\makeatletter
\newcommand{\@BIBLABEL}{\@emptybiblabel}
\newcommand{\@emptybiblabel}[1]{}
\makeatother
\usepackage[hidelinks]{hyperref}
\makeatletter
\newcommand\figcaption{\def\@captype{figure}\caption}
\newcommand\tabcaption{\def\@captype{table}\caption}
\makeatother
\title{Leveraging Term Banks for Answering Complex Questions: \\
  A Case for Sparse Vectors}

\author{Peter D. Turney \\
  Allen Institute for Artificial Intelligence \\
  2157 N Northlake Way Suite 110, Seattle, WA 98103 \\
  {\tt peter.turney@gmail.com}}

\date{}

\begin{document}

%
%
\setlength{\abovedisplayskip}{-6pt}
\setlength{\belowdisplayskip}{8pt}

\maketitle

\begin{abstract}
While open-domain question answering (QA) systems have proven effective for answering simple questions, 
they struggle with more complex questions. Our goal is to answer more complex questions reliably, 
without incurring a significant cost in knowledge resource construction to support the QA. One readily 
available knowledge resource is a {\em term bank}, enumerating the key concepts in a domain. We have 
developed an unsupervised learning approach that leverages a term bank to guide a QA system, by 
representing the terminological knowledge with thousands of specialized vector spaces. In experiments 
with complex science questions, we show that this approach significantly outperforms several 
state-of-the-art QA systems, demonstrating that significant leverage can be gained from continuous 
vector representations of domain terminology.

\hspace{4pt} In our experiments, we made the surprising discovery that dense, low-dimensional 
embeddings (used in many AI systems) were not the most effective representation, and that
sparse, high-dimensional vector spaces performed better. We discuss the reasons for this, and the 
implications this may have for other projects that have assumed embeddings are the best continuous 
representation.
\end{abstract}

%
%
 
\section{Introduction}

{\em Open-domain} question answering (QA) systems typically use information retrieval (IR) techniques 
to answer questions by searching in a large corpus of natural language text \cite{Strzalkowski:06}.
They support relatively simple queries, such as questions about facts involving named entities. More 
complex queries are possible with {\em restricted-domain} QA systems \cite{Molla:07}. These systems 
generally use classical AI techniques, such as rule-based systems with knowledge bases. 

We aim to answer complex questions in a restricted domain without the use of knowledge bases or other
expensive resources. Our chosen domain is science at the levels of elementary school (3rd to 5th grade) 
and middle school (6th to 8th grade). We use multiple-choice science exam questions to evaluate our QA 
system. Figure~\ref{fig:qexample} shows an example of a middle school exam question.

\begin{table}[ht]
\small
\begin{center}
\begin{tabular}{|l|}
\hline
Which of the following statements best explains why \\
earthquakes occur more frequently in California \\
than in Massachusetts? \\
(A) The rock found in California is igneous, but the \\
\hspace{14pt} rock found in Massachusetts is sedimentary. \\
(B) California is located on the boundary of two \\
\hspace{14pt} crustal plates, but Massachusetts is not. \\
(C) The rock under California is soft, but the rock \\
\hspace{14pt} under Massachusetts is hard. \\
(D) California is located on a continental plate, but \\
\hspace{14pt} Massachusetts is not. \\
\hline
\end{tabular}
\end{center}
\normalsize
\vspace{-6pt}
\figcaption{\label{fig:qexample} A middle school multiple-choice exam question. 
The correct answer is (B).}
\end{table}

Several recent papers address answering multiple-choice science exam questions 
\cite{Khot:15,Clark:16,Jauhar:16,Khashabi:16}. Multiple-choice exams are an excellent benchmark for 
QA systems, since the questions are complex, yet performance is easily measured. 

Our approach to restricted-domain QA is to assume that the domain will have a specific vocabulary, 
in the form of a term bank, which can guide the QA system. The intuition is, for every 
question, there is a key concept that links the question to the best answer. If we can identify the term 
that expresses the key concept, then we have an excellent guide to finding the correct answer. This 
intuition is related to {\em lexical cohesion} in discourse. Morris and Hirst~\shortcite{Morris:91} 
describe lexical cohesion as ``the cohesion that arises from semantic relationships between words'', 
resulting from ``chains of related words that contribute to the continuity of lexical meaning.'' 

We use a term bank to find a cohesive link between a question and a candidate answer.
For each candidate answer, we search for the term that provides maximal lexical cohesion between 
the question and the answer. The best candidate is the choice with the highest lexical cohesion 
with the question.

Our QA system, {\em Multivex}, uses an unsupervised method to build three types of vector 
spaces: terminology space, word space, and sentence space. {\em Terminology space} is designed for finding a term 
in the term bank that links a question to a candidate answer with strong lexical cohesion. 
{\em Word space} is designed to characterize a word by the context in which the word appears. 
{\em Sentence space} is designed to characterize a sentence by the words that it contains. 

There is only one terminology space, which contains one vector for each term in the term bank. 
There are thousands of word spaces and sentence spaces, one for each term in the term bank. 
The vector representation of a word or a sentence is modulated by the term bank. A word or 
sentence has no global vector representation; it only has a representation with respect to a 
given term. 

For example, consider Figure~\ref{fig:qexample}. Terminology space tells us that the term 
{\em earthquakes} has a high lexical cohesion with the question and answer (B). The word space for
{\em earthquakes} tells us that the word {\em plates} often appears in the context {\em crustal}.
The sentence space for {\em earthquakes} tells us that the question as a whole is similar
to the kinds of sentences that occur in the {\em earthquakes} sentence space. The three spaces 
all agree that there is a high lexical cohesion between the question and answer (B). 

In our prototypes of Multivex, at first we used dense, low-dimensional embeddings for our vector 
spaces, since embeddings have achieved impressive results on a variety of tasks \cite{Mikolov:13a}. 
We were surprised to later find that sparse, high-dimensional vector spaces yielded better results.

The reason sparse vectors work well in our QA system is that rare word co-occurrences provide the 
strongest evidence for lexical cohesion. When a term such as {\em earthquakes} links {\em plates} 
and {\em crustal}, this is a rare event that signals an important connection. The problem with dense, 
low-dimensional embeddings is that they smooth away rare events. Dense embeddings are good 
for capturing the general usage of a word such as {\em plates}, but they ignore specialized word
senses, such as {\em crustal plates}.

We have two main results: (1) Leveraging term banks is an inexpensive way to answer complex questions.
Term banks are a good source for the concepts that make an answer lexically cohesive with a 
question. (2) Sparse vectors capture lexical cohesion better than dense vectors. Dense vectors are good
for capturing the general sense of a word, but facts lie at the intersection of several word meanings;
facts tend to be rare and specific, which makes sparse vectors more appropriate when seeking facts.

In the following section, we discuss related work with science exam questions and past analysis
of sparsity versus density. Section~\ref{sec:multivex} presents a detailed description of Multivex. 
In Section~\ref{sec:experiments}, we show that Multivex performs better on science exam questions 
than a strong IR baseline. We compare sparse vectors to Word2vec embeddings \cite{Mikolov:13b} and 
truncated singular value decomposition (SVD) embeddings \cite{Turney:10}, demonstrating that Multivex 
works best with sparse vectors. Section~\ref{sec:discussion} discusses the results of the experiments. 
We consider limitations of Multivex in Section~\ref{sec:future} and we conclude in 
Section~\ref{sec:conclusion}. 

%
%

\section{Related Work}

The first TREC (Text REtrieval Conference) QA task took place in 1999 \cite{Voorhees:99}.
The task was to answer fact-based, short-answer, open-domain questions, mostly involving named 
entities, by retrieving small snippets of text. We now have robust, well-tested IR techniques 
for answering these kinds of questions and research is shifting to more challenging problems. 

In this section, we discuss related work with science exam questions. Since embeddings are
currently popular, our results with sparse vectors may be surprising; therefore we also discuss
past work that compares sparse vectors to embedddings.

\subsection{Multiple-Choice Science Exam Questions}

Past work with science exam questions has used structured information, in the form of if-then
rules or tables. This information tends to be unreliable if it is acquired automatically
or labor-intensive if it is acquired manually. Multivex needs only a large corpus of text and a 
term bank for the chosen domain. 

Khot et al.~\shortcite{Khot:15} compared three different types of Markov Logic Networks (MLNs)
for answering science exam questions. They used structured knowledge in the form of if-then 
rules.

Clark et al.~\shortcite{Clark:16} evaluated an ensemble of five solvers: three of the five
were corpus-based, but the fourth used if-then rules and the fifth used tables. Their ablation 
study demonstrated that all five solvers made a significant contribution. 

Jauhar et al.~\shortcite{Jauhar:16} represented science knowledge in a tabular form, where
rows stated facts and columns imposed a parallel structure of types on the rows. The
best answer to a question was determined by the row and column that best supported
one of the choices. They trained a supervised log-linear model to score the choices.

Khashabi et al.~\shortcite{Khashabi:16} applied ILP to knowledge in a tabular form, using the
same tables as Jauhar et al.~\shortcite{Jauhar:16}. Their ILP system performed multi-step inference
by chaining together multiple rows from separate tables. 

\subsection{Sparsity and Density}

Dense embeddings achieve good results on many tasks \cite{Turney:10}. The classical approach to 
embeddings is to make a word--context co-occurrence matrix and then apply a dimensionality 
reduction algorithm \cite{Landauer:97}. A more recent approach is to learn embeddings with a 
neural network \cite{Mikolov:13a,Mikolov:13b}. Baroni et al.~\shortcite{Baroni:14} described the 
classical approach as {\em context-counting} and the neural network approach as 
{\em context-predicting}. However, Levy et al.~\shortcite{Levy:14b} argued the two approaches 
are learning the same latent structure.

Many papers report that dense embeddings are better than sparse vectors. For example, Landauer 
et al.~\shortcite{Landauer:97} achieved 64.4\% on the TOEFL synonym test with embeddings from 
truncated SVD, but the original sparse matrix only achieves 36.8\%. 

In a series of papers, Levy et al.~\shortcite{Levy:14a,Levy:14b,Levy:15} compared sparse and dense 
vectors. In summary, they reported that ``there is no single method that consistently performs 
better than the rest'' \cite{Levy:15} and a sparse representation ``is superior in some of the 
more semantic tasks'' \cite{Levy:14a}. Toutanova et al.~\shortcite{Toutanova:15} show that a
sparse ``observed features'' model is better than a dense ``latent feature'' model for
knowledge bases and textual inference.

%
%

\section{Multivex}
\label{sec:multivex}

The inputs to Multivex are a term bank, a corpus, and a multiple-choice question. The output is 
the answer to the question. Multivex uses three types of spaces: terminology space, word space, 
and sentence space. Each term in the term bank maps to one row vector in the terminology matrix, 
one word matrix in the set of word matrices, and one sentence matrix in the set of sentence matrices. 
Table~\ref{tab:spaces} summarizes the spaces.

\begin{table*}[ht]
\small
\begin{center}
\begin{tabular}{lll}
\hline
Matrices (Spaces) & Rows (Entities) & Columns (Features) \\
\hline
1 terminology matrix    & 9,009 science terms                      & 22,767,476 unigrams and conjunctions \\
9,009 word matrices     & 2,081 words on average per matrix        & millions of unigrams, bigrams, and trigrams \\
9,009 sentence matrices & 16,155 sentences on average per matrix   & millions of unigrams, bigrams, and trigrams \\
\hline
\end{tabular}
\end{center}
\normalsize
\vspace{-6pt}
\caption{\label{tab:spaces} A summary of the three types of spaces.}
\end{table*}

Let $q$ be a question with $m$ possible answers \mbox{$A = \{ a_1, \ldots, a_m \}$} and let
$T = \{ t_1, \ldots, t_n \}$ be our term bank with $n$ terms. Multivex scores each QA pair
$\langle q, a_i \rangle$ with respect to a science term $t_j$. The score $\textit{score}(q, a_i | t_j)$
is an average over eight subscores, four based on terminology space, two based on word space, and two 
based on sentence space. The final score for the pair $\langle q, a_i \rangle$ is the maximum 
score over all $t_j \in T$. The best guess for the correct answer to $q$ is the answer $a_*$ with 
the highest score over all $a_i \in A$ and all $t_j \in T$. 

To construct these vector spaces for a given domain, we begin with a term bank for the domain and a 
large corpus of text, such that most of the terms in the term bank occur frequently in the corpus. 
We then build a set of {\em pseudo-documents}, one for each term, by taking the union of the sentences 
in the corpus that contain the given term. From the pseudo-documents, we build terminology space, word 
space, and sentence space. These three types of spaces are used to calculate $\textit{score}(q, a_i | t_j)$.

\subsection{Term Bank}

In our case, the given domain is elementary and middle school science. The term bank consists of 
9,356 terms from 52 science glossaries.\footnote{\scriptsize \url{http://allenai.org/data.html}}
Most of the glossaries came from \mbox{K-12} (kindergarten and 1st to 12th grades) websites. 

\subsection{Corpus}
\label{subsec:corpus}

The corpus consists of 280 GB of text (50 billion tokens) collected by a web crawler. 
All markup was removed from the web pages and the text was split into sentences with the Stanford CoreNLP 
sentence segmenter.\footnote{\scriptsize \url{http://stanfordnlp.github.io/CoreNLP/}} We selected
English text by requiring all sentences to contain English stop words, using the SMART stop 
word list \cite{Salton:71}.\footnote{\scriptsize \url{http://www.lextek.com/manuals/onix/stopwords2.html}} 
The result was 1.75 billion English sentences. 

\subsection{Pseudo-Documents}

For each of the 9,356 science terms, we searched in the corpus for sentences that contained the given term. 
If there were fewer than ten sentences, we dropped the term, leaving us with 9,009 science terms. For 
each remaining term, we collected a maximum of 50,000 sentences, which formed the pseudo-document for the 
term.\footnote{\scriptsize \url{http://allenai.org/data.html}} 

\subsection{Terminology Space}
\label{subsec:terminology}

Terminology space is designed to find the best linking concept (the best science term) for a given QA pair. 
Terminology space consists of a single matrix with 9,009 rows, one row for each science term. Each
row is a sparse vector with 22,767,476 columns, where the columns are features derived from the
sentences in the pseudo-documents for the science terms. Each sentence was converted into a set of features,
consisting of unigrams and conjunctions of unigrams. Sentences were tokenized with the Stanford CoreNLP
tokenizer and tokens were stemmed and converted to lower case. 

The motivation for the conjunction features is to represent potential cohesive links. For example, 
consider Figure~\ref{fig:qexample}. In the row vector for {\em earthquakes}, the conjunction feature 
{\em boundary \& earthquake} has a high tf--idf (term frequency--inverse document frequency) weight 
\cite{Salton:88}, and it links the word {\em earthquakes} in the question to the word {\em boundary} in 
the correct answer (B).

For a given row in the terminology matrix (corresponding to a unique science term and a unique 
pseudo-document), any unigram that occurred in ten or more sentences and did not appear in the SMART 
stop word list was selected as a feature for that row. Any two distinct unigrams (excluding stop words) 
that occurred together in the same sentence, in a window of ten words, in ten or more sentences were 
selected as a conjunction feature for that row. Conjunction features were normalized by putting the 
component unigrams in alphabetical order. Table~\ref{tab:features} shows the top features for the  
term {\em earthquakes}.

\begin{table}[ht]
\small
\begin{center}
\begin{tabular}{rl}
\hline
Frequency & Feature \\
\hline
49,944 & earthquake \\
4,149 & flood \\
4,064 & earthquake \& flood \\
3,709 & volcano \\
3,604 & earthquake \& volcano \\
3,254 & earth \\
3,117 & occur \\
3,062 & earthquake \& occur \\
2,969 & natural \\
2,936 & disaster \\
\hline
\end{tabular}
\end{center}
\normalsize
\vspace{-6pt}
\caption{\label{tab:features} The top features for {\em earthquakes}.}
\end{table}

We convert the raw frequency counts into tf--idf  values and binary values. Consider the conjunction 
feature {\em earthquake \& flood} for the science term {\em earthquakes}. Suppose {\em earthquakes} 
corresponds to the $i$-th row in the terminology matrix and {\em earthquake \& flood} corresponds to 
the $j$-th column. The term frequency $\mathit{tf_{ij}}$ is the number of sentences in the pseudo-document 
for {\em earthquakes} that contain {\em earthquake \& flood}; that is, $\mathit{tf_{ij}} = \textrm{4,064}$ 
(see Table~\ref{tab:features}). The document frequency $\mathit{df_j}$ is the number of pseudo-documents 
for which the feature is nonzero. The tf--idf weight $w_{ij}$ for the feature {\em earthquake \& flood} 
in the science term {\em earthquakes} is defined as follows: 

\begin{align}
\mathit{TF_{ij}} &= \frac{\log_{10} ( \mathit{tf_{ij}} + 1 )}{\max_j \log_{10} ( \mathit{tf_{ij}} + 1 )} \\
\mathit{IDF_j} &= 1 - \frac{\log_{10} ( \mathit{df_j} + 1 )}{\max_j \log_{10} ( \mathit{df_j} + 1 )} \\
w_{ij} &= \mathit{TF_{ij}} \cdot \mathit{IDF_j}
\label{eq:wij}
\end{align}

\noindent The tf--idf weight $w_{ij}$ ranges between 0 and 1. The binary weight is zero if
the tf--idf weight is zero; otherwise, it is one.

\subsection{Word Space}

Word space is designed to characterize how a word behaves in the context of a given scientific
term. For example, the context that surrounds the word {\em boundary} in sentences about
{\em earthquakes} will be different from the context around {\em boundary} in sentences about
{\em solid state}. The idea is to evaluate whether the words in a QA pair are being used in
the QA pair in the same sense as they are used with the given scientific term. That is, the contexts
in the QA pair should be similar to the contexts in the pseudo-document for the scientific term.
If they are not similar, then the term is not a good match for the QA pair. 

This can be viewed as a kind of word sense disambiguation. The vector representation of {\em boundary} 
is modulated by the scientific terms {\em earthquakes} and {\em solid state}. By choosing the
term, we choose the sense of {\em boundary} \cite{Reisinger:10}.

There are 9,009 word space matrices, one for each science term. The word matrix for a given science
term is generated from the corresponding pseudo-document for the term. The rows in the word matrix
correspond to all of the unigrams (excluding stop words) that occur ten or more times in the 
pseudo-document. For example, the word matrix for {\em earthquakes} has 5,385 rows, corresponding to
5,385 unique unigrams. The columns in the word matrix are features derived from the contexts around 
the words in the pseudo-document. 

For a given row in a word matrix, the context for the corresponding word (unigram) is defined as
all unigrams, bigrams, and trigrams that appear in a window of three words
to the left and three words to the right of the given word, in all of the sentences in the given 
pseudo-document. The term frequency $\mathit{tf_{ij}}$ for a contextual n-gram feature is the number 
of tokens of the given word, such that the n-gram occurs in the context of the token. The document 
frequency $\mathit{df_j}$ is the number of words such that the n-gram appears in some context 
of the word. The weight $w_{ij}$ is defined as in Equation~\ref{eq:wij}.

\subsection{Sentence Space}
\label{subsec:sentence}

Sentence space is intended to model the typical sentences that contain the given science term.
The aim is to treat the given QA pair as if it were a sentence, and then compare it to the
sentences in sentence space. If the given scientific term is appropriate for the given QA pair,
then the QA pair should be similar to sentences in the pseudo-document for the scientific term.
For instance, one of the sentences in the pseudo-document for {\em earthquakes} is, ``For example, 
major earthquakes regularly occur along California's San Andreas fault -- a giant fracture in the 
Earth that marks the boundary between the North American and Pacific tectonic plates.''
Compare this sentence to the question in Figure~\ref{fig:qexample}. In this example, the sentence
covers the QA pair thoroughly, but we do not assume that a single sentence will contain all of 
the information that we need to answer a question. Sentence space is used to calculate subscores
that combine information from several parts of several sentences (see Section~\ref{subsec:scoring}).

There are 9,009 sentence space matrices, one for each science term. The sentence matrix for a
given science term is generated from the corresponding pseudo-document for the term. The rows
correspond to all of the sentences in the pseudo-document. The columns correspond to all of 
the unigrams, bigrams, and trigrams in the pseudo-document. The sentence matrix for {\em earthquakes} 
has 50,000 rows, corresponding to the 50,000 sentences that appear in the pseudo-document for 
{\em earthquakes}.

For the sentence matrices, we found that binary vectors worked better than tf--idf weighted
vectors. In other words, a sentence row vector is simply the set of n-grams that appear in the
sentence.

\subsection{Scoring QA Pairs}
\label{subsec:scoring}

The score, $\textit{score}(q, a_i | t_j)$, for a QA pair, $\langle q, a_i \rangle$, is the average of 
eight subscores that are calculated in four steps, two subscores per step. All of the subscores are 
designed to measure the {\em lexical cohesion} between the question and the candidate answer. 

The subscores are weighted inner products of vectors. They all range from 0 to 1. To answer questions 
quickly, we do not calculate all subscores for every scientific term $t_j \in T$. As we go through 
each step, we only advance the best terms to the next step. 

{\bf Step~1:} {\em terminology space with tf--idf weights:} In this step, we calculate two subscores
using the terminology matrix. We iterate over all 9,009 scientific terms, searching for the top ten 
terms that maximize the average of the first two subscores. The QA pair must be converted into 
unigrams and conjunctions, so that it can be compared to the row vectors in terminology space. We first 
process the question and the answer separately, generating unigrams and conjunctions for each as if 
they were two separate sentences. We then create further conjunction features by pairing every unigram 
in the question $q$ with every unigram in the answer $a_i$. 

{\bf Step~1.1:} {\em tf-idf weighted unigram match:} Let $\mathbf{v}_{ru}(t_j)$ be the sparse tf--idf 
row vector in terminolgy space that corresponds to the science term $t_j$, where the features are only 
unigrams; all conjunction features are dropped ($ru$ for {\em real-valued unigrams}). Let 
$\mathbf{v}_{bu}(q, a_i)$ be the sparse binary vector that represents the QA pair, where the features 
are only unigrams ($bu$ for {\em binary unigrams)}. We define the subscore for Step~1.1 as follows:

\begin{equation}
\textit{score}_{1.1}(q, a_i | t_j) = 
\frac{\mathbf{v}_{ru}(t_j) \cdot \mathbf{v}_{bu}(q, a_i)}{|\mathbf{v}_{bu}(q, a_i)|_1}
\label{eq:score1.1}
\end{equation}

\noindent Here $|\mathbf{x}|_1$ is the L1 norm of the vector $\mathbf{x}$ and $\mathbf{x} \cdot \mathbf{y}$ 
is the inner product of the vectors $\mathbf{x}$ and $\mathbf{y}$. 

{\bf Step~1.2:} {\em tf-idf weighted conjunction match:} Let $\mathbf{v}_{rc}(t_j)$ be the sparse tf--idf 
row vector in terminology space that corresponds to the science term $t_j$, where the features are only 
conjunctions; all unigram features are dropped ($rc$ for {\em real-valued conjunctions}). Let 
$\mathbf{v}_{bc}(q, a_i)$ be the sparse binary vector that represents the QA pair, where the features 
are only conjunctions ($bc$ for {\em binary conjunctions}). We define the subscore for Step~1.2 as follows:

\begin{equation}
\textit{score}_{1.2}(q, a_i | t_j) = 
\frac{\mathbf{v}_{rc}(t_j) \cdot \mathbf{v}_{bc}(q, a_i)}{|\mathbf{v}_{bc}(q, a_i)|_1}
\label{eq:score1.2}
\end{equation}

{\bf Step~2:} {\em terminology space with binary weights:} In this step, we only iterate over the
top ten science terms that maximize the average of the two subscores from Step 1. We use terminology space
again, but we convert the science term feature weights from tf--idf to binary. 

{\bf Step~2.1:} {\em binary unigram match:} This is the same as Step~1.1, except with binary term weights:

\begin{equation}
\textit{score}_{2.1}(q, a_i | t_j) = 
\frac{\mathbf{v}_{bu}(t_j) \cdot \mathbf{v}_{bu}(q, a_i)}{|\mathbf{v}_{bu}(q, a_i)|_1}
\end{equation}

{\bf Step~2.2:} {\em binary conjunction match:} This is the same as Step~1.2, except with binary term weights:

\begin{equation}
\textit{score}_{2.2}(q, a_i | t_j) = 
\frac{\mathbf{v}_{bc}(t_j) \cdot \mathbf{v}_{bc}(q, a_i)}{|\mathbf{v}_{bc}(q, a_i)|_1}
\end{equation}

{\bf Step~3:} {\em word space with tf--idf weights:} In this step, we only iterate over the top four 
science terms that maximize the average of the four subscores from Steps 1 and 2. We use word space to
calculate two subscores. For each QA pair, we consider four science terms, corresponding to four word
matrices. To compare the QA pair to a word matrix, the pair must be converted into words and for each QA 
word we need to find the corresponding context, which is the set of unigrams, bigrams, and trigrams that 
occur in a window of three words to the left and three words to the right of the given word. We compare 
the contexts in the QA pair to the contexts in the word spaces.

{\bf Step~3.1:} {\em word context match with same word:} Let $w$ be a word in the QA pair. This subscore 
measures the degree to which the context of $w$ in the QA pair is similar to the  context of $w$ in the 
word space for $t_j$, for an average $w$. Let $\mathbf{v}_{rw}(w | t_j)$ be the sparse tf--idf context 
vector that represents $w$ in the word matrix for $t_j$ ($rw$ for {\em real-valued words}) and let 
$\mathbf{v}_{bw}(w | q, a_i)$ be the sparse binary context vector that represents $w$ in the QA pair 
($bw$ for {\em binary words}). In both vectors, $\mathbf{v}_{rw}(w | t_j)$ and $\mathbf{v}_{bw}(w | q, a_i)$, 
the features are unigrams, bigrams, and trigrams that occur in a window around $w$. The score for $w$ 
is defined as follows:

\begin{equation}
\! \! \textit{score}_w(w | q, a_i, t_j) \! = \!
\frac{\mathbf{v}_{rw}(w | t_j) \cdot \mathbf{v}_{bw}(w | q, a_i)}{|\mathbf{v}_{bw}(w | q, a_i)|_1}
\end{equation}

\noindent If $w$ does not correspond to a row in the word matrix for the term $t_j$, then 
$\textit{score}_w(w | q, a_i, t_j)$ is zero. Let $W$ be the set of all words (unigrams excluding stop words)
in the QA pair. The subscore for Step~3.1 is defined as the average of the word scores:

\begin{equation}
\textit{score}_{3.1}(q, a_i | t_j) = 
\operatorname*{avg}\limits_{w \in W}{\textit{score}_w(w | q, a_i, t_j)}
\end{equation}

{\bf Step~3.2:} {\em word context match with different words:} This subscore measures the degree to 
which, for the average word $x$ in $q$ (or in $a_i$), there is a word $y$ in $a_i$ (or in $q$) such 
that the context around $x$ in the QA pair is similar to the context around $y$ in the word matrix that 
corresponds to the science term $t_j$. Let $x$ and $y$ be words in the QA pair, such that one of them is 
from $q$ and the other is from $a_i$. For each $x$, we want to find the word $y$ that has the most similar 
context with $x$; we are looking for context that connects the question $q$ to the answer $a_i$. Let 
$\mathbf{v}_{bw}(x | q, a_i)$ be the sparse binary context vector that represents $x$ and let 
$\mathbf{v}_{rw}(y | t_j)$ be the sparse tf--idf context vector that represents $y$ in the word 
matrix for $t_j$. If $x$ is in $q$, let $Y$ be the set of words in $a_i$; otherwise, if $x$ is in $a_i$, 
let $Y$ be the set of words in $q$. The score for $x$ is defined as follows:

\begin{multline}
\textit{score}_{xy}(x | q, a_i, t_j) = \\
\operatorname*{max}\limits_{y \in Y}{
\frac{\mathbf{v}_{rw}(y | t_j) \cdot \mathbf{v}_{bw}(x | q, a_i)}{|\mathbf{v}_{bw}(x | q, a_i)|_1}
}
\end{multline}

\noindent Let $W$ be the set of all words in the QA pair. The subscore is the average of the word scores:

\begin{equation}
\textit{score}_{3.2}(q, a_i | t_j) = 
\operatorname*{avg}\limits_{w \in W}{\textit{score}_{xy}(w | q, a_i, t_j)}
\end{equation}

{\bf Step~4:} {\em sentence space with binary weights:} In this step, we only consider the first
top science term that maximizes the average of the six subscores from Steps 1, 2, and 3. We use
sentence space to calculate two subscores. For each QA pair, we only evaluate the one sentence
matrix that corresponds to the top science term. We treat the QA pair as if it were a sentence and
we compare it to the sentences in the sentence matrix. To compare the QA pair to a row vector in
the sentence matrix, we extract all unigrams, bigrams, and trigrams from $q$ and from $a_i$.

{\bf Step~4.1:} {\em sentence whole match:} Let $s$ be a sentence in the set of sentences $S$ in 
the pseudo-document that corresponds to the term $t_j$. Let $\mathbf{v}_{bs}(s | t_j)$ be the sparse 
binary row vector in the sentence matrix for $t_j$ that corresponds to the sentence $s \in S$ ($bs$ for 
{\em binary sentence}). We also represent the QA pair as if it were a sentence. Let 
$\mathbf{v}_{bs}(q, a_i)$ be a sparse binary vector in which the features are all unigrams, bigrams, 
and trigrams from $q$ and from $a_i$. We score $s \in S$ by its similarity to the QA pair:

\begin{equation}
\textit{score}_{sw}(s | q, a_i, t_j) = 
\frac{\mathbf{v}_{bs}(s | t_j) \cdot \mathbf{v}_{bs}(q, a_i)}{|\mathbf{v}_{bs}(q, a_i)|_1}
\end{equation}

\noindent Let $S_k$ be the top $k$ sentences in $S$ that have the highest similarity scores
for the given QA pair (we set $k$ to five). The subscore is the average of the $k$ scores:

\begin{equation}
\textit{score}_{4.1}(q, a_i | t_j) = 
\operatorname*{avg}\limits_{s \in S_k}{\textit{score}_{sw}(s | q, a_i, t_j)}
\end{equation}

{\bf Step~4.2:} {\em sentence subset match:} For this subscore, we search for a subset
of the words in the QA pair such that the context around that subset has a large overlap with
a sentence in $S$ for $t_j$. Let $U_m$ be the set of all subsets of unigrams in the QA pair, 
up to a maximum of $m$ unigrams per subset (we set $m$ to six). For $u \in U_m$, let $c(u)$
be the union of the contexts for every unigram in $u$. Thus $c(u)$ contains the unigrams, bigrams, 
and trigrams from $q$ and from $a_i$ that occur three words to the left and right of each unigram 
in $u$. Let $\mathbf{v}_{bs}(c(u) | q, a_i)$ be a sparse binary vector of the n-grams in $c(u)$. 
Let $\mathbf{v}_{bs}(s | t_j)$ be the sparse binary row vector in the sentence matrix for $t_j$ 
that corresponds to the sentence $s \in S$. Let $|u|$ be the cardinality of the set $u$. We score 
the sentence $s$ by its maximum similarity to $c(u)$:

\begin{multline}
\textit{score}_{ss}(s | q, a_i, t_j) = \\
\operatorname*{max}\limits_{u \in U_m}
\left(
\frac{\mathbf{v}_{bs}(s | t_j) \cdot \mathbf{v}_{bs}(c(u) | q, a_i)}{|\mathbf{v}_{bs}(c(u) | q, a_i)|_1}
\cdot \frac{|u|}{m}
\right)
\end{multline}

\noindent Let $S_k$ be the top $k$ sentences in $S$ that have the highest similarity scores
for the given QA pair (we set $k$ to five). The subscore is the average of the $k$ scores:

\begin{equation}
\textit{score}_{4.2}(q, a_i | t_j) = 
\operatorname*{avg}\limits_{s \in S_k}{\textit{score}_{ss}(s | q, a_i, t_j)}
\end{equation}

The final score for the QA pair, $\langle q, a_i \rangle$, is the average of the eight subscores
above, given the top science term $t_j$ selected by the four steps. 

%
%

\section{Experiments}
\label{sec:experiments}

In this section, we first test whether Multivex can surpass an IR baseline system. We then 
replace some of the sparse, high-dimensional vectors in Multivex with dense, low-dimensional 
vectors, to determine the effect of dimensionality and density. We also measure how much
each of the eight subscores is contributing to the accuracy and how sensitive Multivex is 
to parameter settings. 

In the following experiments, we use multiple-choice science exam questions at the elementary 
school (3rd to 5th grade) and middle school (6th to 8th grade) levels.\footnote{\scriptsize 
\url{http://allenai.org/data.html}.} The questions have been divided into train, 
development, and test subsets, summarized in Table~\ref{tab:qsets}. We used the train and development
subsets while developing Multivex and we used the test subsets for the experiments that follow.

\begin{table}[ht]
\small
\begin{center}
\begin{tabular}{lrrrr}
\hline
Questions           & Train & Dev & Test & Total \\
\hline
Public Elementary   & 432  &  84 &  339 &  855 \\
Public Middle       & 293  &  65 &  282 &  640 \\
Licensed Elementary & 574  & 143 &  717 & 1434 \\
Licensed Middle     & 1581 & 482 & 1631 & 3694 \\
\hline
All Questions       & 2880 & 774 & 2969 & 6623 \\
\hline
\end{tabular}
\end{center}
\normalsize
\vspace{-6pt}
\caption{\label{tab:qsets} Number of questions in each set.}
\end{table}

\subsection{Comparison with an IR Baseline}
\label{subsec:lucene}

In their experiments with multiple-choice science exam questions, Clark et al.~\shortcite{Clark:16} 
show that a QA system based on Lucene\footnote{\scriptsize \url{https://lucene.apache.org/}} is a strong 
approach to question answering, out-performing several more complex algorithms. Therefore our first 
experiment compares Multivex with a Lucene-based approach.

Table~\ref{tab:algorithms} shows the accuracy of Lucene and Multivex on the test questions.
The algorithms assign a score to each of the four possible answer choices and the choice
with the highest score is selected as the best guess. Accuracy is measured by the percentage 
of correct choices. If $n$ answer choices tie for the correct score, the algorithm gets
a partial mark of $1/n$, the expected value of randomly resolving ties. Both algorithms use
the same corpus of 1.75 billion English sentences, described in Section~\ref{subsec:corpus}.

\begin{table*}[ht]
\small
\begin{center}
\begin{tabular}{clccccccr}
\hline
Section & Algorithm  & Public     & Public & Licensed   & Licensed & All Test  & Vector & Delta \\
        &            & Elementary & Middle & Elementary & Middle   & Questions & Type \\
\hline
4.1 & Multivex   & 59.7 & 60.6 & 51.1 & 49.0 & 51.8 & sparse &  $0.0$ \\
4.1 & Lucene     & 55.8 & 52.5 & 48.7 & 47.3 & 49.1 & sparse & $-2.7$ \\
\hline
4.2 & SVD 1      & 55.5 & 51.5 & 46.8 & 45.6 & 47.6 & dense  & $-4.3$ \\
4.2 & SVD 2      & 56.2 & 51.8 & 48.3 & 46.9 & 48.8 & dense  & $-3.1$ \\
\hline
4.3 & Word2vec 1 & 49.9 & 49.7 & 41.7 & 42.2 & 43.7 & dense  & $-8.2$ \\
4.3 & Word2vec 2 & 51.9 & 52.6 & 45.3 & 44.8 & 46.5 & dense  & $-5.4$ \\
\hline
\end{tabular}
\end{center}
\normalsize
\vspace{-6pt}
\caption{\label{tab:algorithms} Accuracy of various algorithms on the test question sets.
Delta is the accuracy of an algorithm on all test questions minus the accuracy of Multivex.}
\end{table*}

Multivex achieved higher scores than Lucene on all test question sets. The difference between Multivex
(51.8\%) and Lucene (49.1\%) is statistically significant, using Fisher's Exact test with a significance 
level of 95\%. For comparison with past work, Jauhar et al.~\shortcite{Jauhar:16} achieve 54.9\% on the 
Public Elementary questions, but this result is for the whole set of 855 questions, not the test set. 
Sachan et al.~\shortcite{Sachan:16} achieve 46.7\% on the whole Public Elementary set (855 questions). 
Baudis et al.~\shortcite{Baudis:16} achieve 44.1\% on the Public Middle test set (282 questions).

\subsection{SVD Embeddings}
\label{subsec:svd}

To compare sparse vectors with dense vectors, we modify the first two subscores 
(Steps 1.1 and 1.2) to use dense, low-dimensional SVD embeddings \cite{Landauer:97,Turney:10}. 
We focus on these two subscores because Step 1 plays a key role in Multivex: searching through the 
9,009 scientific terms to pick out the top ten. We leave the other subscores as they are, to make 
it easier to interpret the results.

The embeddings are generated by SVDLIBC,\footnote{\scriptsize \url{https://tedlab.mit.edu/~dr/SVDLIBC/}}
which decomposes the terminology matrix into the product of three matrices $\mathbf{U \Sigma V^\mathsf{T}}$,
where $\mathbf{U}$ and $\mathbf{V}$ are column orthonormal and $\mathbf{\Sigma}$ is a diagonal matrix
of singular values \cite{Golub:96}. The embedding space is given by the matrix $\mathbf{U}_k \mathbf{\Sigma}_k$,
where $\mathbf{U}_k$ is composed of the first $k$ columns of $\mathbf{U}$ and $\mathbf{\Sigma}_k$
is the first $k$ singular values of $\mathbf{\Sigma}$. We use the widely recommended setting $k = 300$.

We evaluate two different ways to apply SVD to terminology space. (1) We simply apply SVD to the
whole terminology matrix. (2) We separate the terminology matrix into two matrices, one matrix with 
all of the unigram features and another matrix with all of the conjunction features. Step 1.1 uses the
first matrix and Step 1.2 uses the second.

Suppose that $\mathbf{e}(t_j)$ is the embedding in $\mathbf{U}_k \mathbf{\Sigma}_k$ that corresponds to the
term $t_j$. In order to compare a QA pair to $\mathbf{e}(t_j)$, we need to project the high-dimensional
vector $\mathbf{v}(q, a_i)$ for the QA pair into the same space as $\mathbf{e}(t_j)$. We can do this by 
multiplying $\mathbf{v}(q, a_i)$ by $\mathbf{V}_k$; that is, $\mathbf{e}(q, a_i) = \mathbf{v}(q, a_i) \mathbf{V}_k$.
Since $\mathbf{v}(q, a_i)$ is sparse, the multiplication is fast.

Table~\ref{tab:algorithms} shows that Multivex with sparse vectors has higher accuracy than
the two SVD variations. The difference between Multivex (51.8\%) and SVD (47.6\% and 48.8\%) is
statistically significant. Lucene also has a higher accuracy (49.1\%) than the two SVD variations, 
but the difference is not statistically significant.

\subsection{Word2vec Embeddings}
\label{subsec:word2vec}

As another comparison of sparse vectors with dense vectors, this experiment evaluates Word2vec 
embeddings \cite{Mikolov:13a,Mikolov:13b} that are pre-trained with Google 
News.\footnote{\scriptsize \url{https://drive.google.com/file/d/0B7XkCwpI5KDYNlNUTTlSS21pQmM/}}
As with SVD, we apply embeddings to the first two subscores (Steps 1.1 and 1.2) and leave the rest 
of Multivex the same. The embeddings are 300-dimensional.

We evaluate two ways to replace the row vectors in terminology space with Word2vec embeddings.
(1) Given a term such as {\em earthquakes}, we simply use the Word2vec vector corresponding to {\em earthquakes}. 
For multi-word terms, we add the Word2vec vectors for each word in the term. (2) Given a term such as 
{\em earthquakes}, we can use the corresponding pseudo-document to make an embedding for {\em earthquakes} 
by taking the sum of the Word2vec embeddings for each unigram in the pseudo-document.

For the unigram match subscore of a QA pair (Step 1.1), we use the average of the cosines 
between the term vector and each vector for unigrams in the QA pair. For 
the conjunction match subscore (Step 1.2), we use the average of the geometric means of the cosines
for each word in the conjunctions. For example, if the term is {\em earthquakes} and the QA pair
has the conjunction {\em flood \& occur}, then we use the geometric mean of
cos({\em earthquakes}, {\em flood}) and cos({\em earthquakes}, {\em occur}).

Table~\ref{tab:algorithms} shows that Multivex with sparse vectors has higher accuracy than
the two Word2vec variations. The difference between Multivex (51.8\%) and Word2vec (43.7\% and 46.5\%) 
is statistically significant. The difference between Lucene (49.1\%) and Word2vec is also
statistically significant.

\subsection{Ablating Subscores}

Table~\ref{tab:ablating} shows the results of ablating each of the eight subscores in Multivex.
All of the subscores make some contribution to the accuracy, except for the two unigram subscores
(Steps 1.1 and 2.1). These subscores appeared to be useful in the training and development sets,
but they are not useful for the test set.

\begin{table}[ht]
\small
\begin{center}
\begin{tabular}{clc}
\hline
Step & Ablated Subscore & Delta \\
\hline
1.1 & tf-idf weighted unigram match      & $+0.2$ \\
1.2 & tf-idf weighted conjunction match  & $-2.1$ \\
2.1 & binary unigram match               & $+0.2$ \\
2.2 & binary conjunction match           & $-1.9$ \\
3.1 & context match with same word       & $-0.6$ \\
3.2 & context match with different words & $-0.8$ \\
4.1 & sentence whole match               & $-1.3$ \\
4.2 & sentence subset match              & $-0.5$ \\
\hline
\end{tabular}
\end{center}
\normalsize
\vspace{-6pt}
\caption{\label{tab:ablating} Ablating subscores in Multivex. 
Delta is the accuracy of the ablated Multivex minus the accuracy of the complete Multivex.
The results are based on the union of all test questions.}
\end{table}

The table suggests that the two conjunction subscores (Steps 1.2 and 2.2) play a key role in Multivex,
based on their deltas. Consider the example in Section~\ref{subsec:terminology} of the conjunction 
feature {\em boundary \& earthquake}. If this feature has a high tf--idf weight, there must be many 
sentences in the pseudo-document for {\em earthquakes} that contain both {\em boundary} and {\em earthquake}. 
This means that there is strong lexical cohesion between the question and answer (B) in 
Figure~\ref{fig:qexample}, given the term {\em earthquakes}.

\subsection{Varying Parameters}

In Section~\ref{subsec:scoring}, we described how Multivex searches through the 9,009 scientific terms.
Step~1 selects the top ten terms, using the first two subscores. Step~2 reduces the ten
down to four. Step~3 picks out the single best term, which is the final output in Step~4. 

These parameter values were tuned on the training sets, with the goal of balancing accuracy and speed.
Table~\ref{tab:varying} explores some alternative values for the parameters, showing the results
on the test set. The default settings, given in Section~\ref{subsec:scoring}, correspond to the
second row in the body of the table.

\begin{table}[ht]
\small
\begin{center}
\begin{tabular}{cccccrcr}
\hline
\multicolumn{4}{c}{Number of Top Terms} & \multicolumn{2}{c}{All Test Qs} \\
\hline
Step 1 & Step 2 & Step 3 & Step 4 & Score & Time \\
\hline
 5 &  2 & 1 & 1 & 51.1 &   2.8 \\
10 &  4 & 1 & 1 & 51.8 &   5.0 \\
20 &  8 & 2 & 1 & 51.9 &  10.4 \\
40 & 16 & 4 & 1 & 51.9 &  20.9 \\
\hline
\end{tabular}
\end{center}
\normalsize
\vspace{-6pt}
\caption{\label{tab:varying} Varying parameters in Multivex. 
The results are based on the union of all test questions.
Time is given in seconds per question.}
\end{table}

The parameter values have little impact on the accuracy (which is a good thing), but they have a
big impact on the execution time (as expected). Answering a single question involves four passes 
through the steps in Section~\ref{subsec:scoring}, one pass for each of the four candidate answers.
With the default parameter settings, Multivex can answer a typical four-choice question in five 
seconds, running on a standard desktop computer (iMac Intel Core i7). 

%
%

\section{Trouble with Embeddings}
\label{sec:discussion}

A problem with embeddings is that rare events tend to be smoothed away. This hypothesis is supported by 
the results in the experiments with SVD and Word2vec (Sections \ref{subsec:svd} and \ref{subsec:word2vec}).

Table~\ref{tab:ablating} shows the value of conjunction features (Steps 1.2 and 2.2). The tf--idf weighted 
conjunction match is the most important subscore. Of the 22,767,476 columns in the terminology matrix, 
22,505,565 are conjunctions (98.8\%). The pseudo-document frequency ($\mathit{df_j}$ in 
Section~\ref{subsec:terminology}) of conjunction features ranges from 1 to 4,292, with a median of 1. 
Conjunction features have a very long tail of rare events. Rare conjunction events convey valuable 
information for answering science questions. 

%
%

\section{Future Work and Limitations}
\label{sec:future}

Our focus in this paper has been multiple-choice questions, but it should be feasible to extend Multivex
to direct-answer questions. For example, the sentence matrices could be used to generate a set of candidate 
direct answers (see Section~\ref{subsec:sentence}).

Multivex is unsupervised; we expect that a supervised approach would yield higher test scores. One possibility 
is to use a supervised deep learning approach with an attention model to focus on rare events 
\cite{Li:16,Zhao:17}.

Another limitation is that the features in Multivex are simple unigrams, bigrams, trigrams, and conjunctions
of unigrams. More complex features, such as part-of-speech tags and semantic relations, could supplement 
these basic features.

The success of our term bank suggests that we should look for other inexpensive resources that can guide 
QA systems. Most of the glossaries that we merged to create our science term bank also contained definitions
for the terms, but we did not use the definitions. A natural improvement to Multivex would be to
exploit the definitions. 

%
%

\section{Conclusion}
\label{sec:conclusion}

Multivex is a restricted-domain QA system, in that it requires a domain-specific term bank, but this 
is a relatively light-weight requirement, compared to QA systems that require if-then rules or tables. 
The key insight is that, with a term bank and some vectors, we can use lexical cohesion to guide us to 
the correct answer.

Multivex is different from much recent work in that it uses sparse, high-dimensional vectors instead of
dense, low-dimensional embeddings. The intuition is that word meanings are distributional and general, but
facts are intersections of word meanings; facts tend to be rare and specific. The experimental results
in Sections \ref{subsec:svd} and \ref{subsec:word2vec} lend support to these intuitions. Replacing sparse 
vectors with dense embeddings reduces the test scores.

As QA systems mature, the emphasis in research will shift from word meanings to sentence meanings. 
Our experience with Multivex suggests that this will require a corresponding shift from dense embeddings
to sparse vectors. Words are repeated but most sentences are unique.

\bibliography{multivex}
\bibliographystyle{acl2012}

\end{document}